# Cluster lens reconstruction using only observed local data

Peter Schneider
Max-Planck-Institut für Astrophysik
Postfach 1523
D-85740 Garching, Germany

## Abstract

The reconstruction of the density profile in clusters of galaxies from the distortion of the images of faint background galaxies is reconsidered. The inversion formula of Kaiser & Squires is known to provide a quantitative way to perform this reconstruction; however, the practical application of this formula faces two problems of principle (besides problems related to the analysis of the observational data): (1) the shear distribution of a lens cannot be inferred from the distortion of images, but only a combination of shear and surface mass density can be observed. (2) The inversion formula is exact only if one assumes observational data on the whole lens plane, whereas in reality, the size of the data field is limited by the size of the CCD. We have considered a possible solution to the first problem in a previous paper. Here we consider the second problem. It is shown that the application of the inversion formula to a finite data field induces systematic boundary effects. An alternative inversion formula is derived, based on some recently published results by Kaiser. We demonstrate, using synthetic data, that this new inversion formula which does not require an extrapolation of the data beyond the observed region, yields results which are comparable with those from the Kaiser & Squires inversion in their 'noise levels', but lack the systemmatic boundary effects.

# 1 Introduction

The determination of the mass distribution in clusters of galaxies from observations of weakly distorted images of faint background galaxies has been recognized as an important tool in observational cosmology. With the pioneering papers of Tyson, Valdes & Wenk (1990) and Kaiser & Squires (1993; henceforth KS; see also Kochanek 1990, Miralda-Escudé 1991), this new method for cluster mass determination has been investigated quantitatively, and several attempts of applying it to real data have been published (Fahlman et al. 1994, Smail, Ellis & Fitchett 1994, Smail et al. 1994). These first applications have demonstrated the great potential of the method, and with the advent of 10 meter class telescopes, the observational situation can be expected to improve dramatically in the next few years. It is therefore of considerable interest to develop and improve the method further. This includes a detailed study of the determination of distortion parameters from observations (e.g., Bonnet & Mellier 1994, Gould 1994), as well as improvements of the underlying inversion technique.

Kaiser & Squires have obtained an exact inversion equation, which yields the surface mass density of the deflector in terms of the shear distribution caused by the lens. Hence, if the shear distribution could be obtained from the observation of distorted images of background sources, the surface mass density of the cluster could be reconstructed. However, the shear is not directly an observable, as was pointed out in Schneider & Seitz (1994, hereafter Paper I); nevertheless, for the case of weak lensing, the observable quantity $g$ (to be defined in Sect. 2 below) is a good approximation to the shear. In Seitz & Schneider (1994, hereafter Paper II), the KS method was generalized to include also the inner part of clusters where the distortion is no longer necessarily weak. In that case, the shear is obtained iteratively from the observables. Note that the strong lensing region of clusters yields particularly strong constraints on the central mass distribution.

In this paper, we want to tackle another problem associated with the KS inversion formula (2.6), namely that it is exact only if the 'data' on the shear are available over the whole lens plane. In practice, however, the finite size of a CCD limits the size of the data field, and in order to apply the inversion formula, an assumption about the shear outside the data field is required. In the above quoted papers the shear was effectively set to zero outside the data field; with this assumption, boundary effects are unavoidable. We have discussed this problem in Paper II where it was shown that these boundary effects can cause artefacts in the reconstructed mass distribution; this is particularly true if the shape of the CCD deviates significantly from that of a square. The 'cure' used in Paper II was an extrapolation of the distortion field to larger distances, which has removed some of these boundary effects; however, such an approach is justified only if the cluster can be considered as an isolated mass distribution. In general, however, this assumption need not be satisfied, and in any case, one should aim for a method which does not make use of any information which is not contained in the observational data. It is therefore necessary to develop an inversion technique which accounts for the finite size of the data field appropriately.

In Sect. 2, the general idea for developing an inversion formula is described, based on a recent paper by Kaiser (1994). The resulting equation is exact on a finite data field and thus has removed the boundary effects. In Sect. 3 the application of this inversion to synthetic galaxy images is briefly described, and in Sect. 4 several examples of this inversion are presented and compared to the results from an KS-like inversion formula.

Our results are summarized and discussed in Sect. 5.

## 2 The inversion method

We use the same notation as in Papers I & II. Hence, for a mass distribution described by its dimensionless surface mass density $\kappa(\boldsymbol{\theta})$, the deflection potential $\psi(\boldsymbol{\theta})$ is defined as

$$\psi(\boldsymbol{\theta}) = \frac{1}{\pi} \int_{\mathbb{R}^2} \mathrm{d}^2\theta' \, \kappa(\boldsymbol{\theta}') \ln \left| \boldsymbol{\theta} - \boldsymbol{\theta}' \right| \quad . \tag{2.1}$$

Since the background sources, which are taken to be very faint galaxies, are much smaller in angular size than the characteristic length scale on which the deflection potential changes, the image of such a faint source can be described in terms of the linearized mapping $\mathrm{d}\boldsymbol{\beta} = A(\boldsymbol{\theta})\,\mathrm{d}\boldsymbol{\theta}$, where $\boldsymbol{\beta}$ denotes the angular position on the source sphere, and

$$A(\boldsymbol{\theta}) = \begin{pmatrix} 1 - \kappa + \gamma_1 & +\gamma_2 \\ +\gamma_2 & 1 - \kappa - \gamma_1 \end{pmatrix} \tag{2.2}$$

is the Jacobian matrix of the lens mapping, which is related to the deflection potential $\psi$ through

$$\begin{aligned} \kappa(\boldsymbol{\theta}) &= \frac{1}{2}\nabla^2 \psi(\boldsymbol{\theta}) \quad , \\ \gamma_1(\boldsymbol{\theta}) &= \frac{1}{2}\left(\psi_{,22} - \psi_{,11}\right) \quad , \\ \gamma_2(\boldsymbol{\theta}) &= -\psi_{,12} \quad , \end{aligned} \tag{2.3}$$

where indices separated by a comma denote partial derivatives with respect to $\theta_i$. For details concerning these lensing relations, cf. Schneider, Ehlers & Falco (1992, henceforth SEF). Combining (2.1 & 3), and defining the complex shear $\gamma(\boldsymbol{\theta}) = \gamma_1(\boldsymbol{\theta}) + \mathrm{i}\gamma_2(\boldsymbol{\theta})$, one obtains

$$\gamma(\boldsymbol{\theta}) = \frac{1}{\pi} \int_{\mathbb{R}^2} \mathrm{d}^2\theta' \, \mathcal{D}(\boldsymbol{\theta} - \boldsymbol{\theta}') \, \kappa(\boldsymbol{\theta}') \quad , \tag{2.4}$$

where

$$\mathcal{D}(\boldsymbol{\theta}) = \frac{\theta_1^2 - \theta_2^2 + 2\mathrm{i}\theta_1\theta_2}{|\boldsymbol{\theta}|^4} \tag{2.5}$$

is a complex kernel. Since (2.4) is a convolution-type integral, its inversion can be most easily performed by using Fourier methods. With these, KS obtained

$$\kappa(\boldsymbol{\theta}) = \frac{1}{\pi} \int_{\mathbb{R}^2} \mathrm{d}^2\theta' \, \mathcal{R}\mathrm{e}\!\left[\mathcal{D}^*(\boldsymbol{\theta} - \boldsymbol{\theta}') \, \gamma(\boldsymbol{\theta}')\right] \quad , \tag{2.6}$$

where $\mathcal{R}\mathrm{e}(x)$ denotes the real part of the complex variable $x$, and the asterisk denotes complex conjugation. Eq. (2.6) is essentially the same as Eq. (2.1.15) of KS, in slightly different notation. Note that the pair of equations (2.4) & (2.6) remains meaningful even for mass distributions for which the deflection potential $\psi$ diverges.[1] Furthermore, note that adding a disk of constant surface mass density does not change $\gamma$ inside the disk; hence, since the data on $\gamma$ are available only in a finite region of the lens plane, the inversion (2.6) is not unique, but determined only up to an additive constant.

---

[1] In order for the integral (2.1) to exist, $\kappa$ must decrease faster than $\theta^{-2}$ if no special symmetries are employed, whereas for the existence of the integrals in (2.4) and (2.6), any decline of $\kappa$ to zero is sufficient.

The problems with (2.6) are that, first, the integral extends over the whole lens plane, whereas the observational data are available on a finite region of the lens plane only, and second, that the shear components are not observable directly. We have dealt with this second problem in detail in Papers I & II. If one assumes that the cluster is not critical, i.e., if the lens does not produce any critical curves, then the quantity

$$g(\boldsymbol{\theta}) := \frac{\gamma(\boldsymbol{\theta})}{1 - \kappa(\boldsymbol{\theta})} \tag{2.7}$$

is an observable (see also Sect. 3). The transformation between source and image ellipticities are such that one cannot distinguish locally between $g$ and $1/g^*$ from image distortions; hence, if one does not assume that the cluster is noncritical, there is a local degeneracy between $g$ and $1/g^*$. For a noncritical cluster, $abs\, g < 1$, and this degeneracy does not occur (see Paper I for more details). By inserting the definition (2.7) into (2.6), we obtain

$$\kappa(\boldsymbol{\theta}) = \frac{1}{\pi} \int_{\mathbb{R}^2} d^2\theta' \left[1 - \kappa(\boldsymbol{\theta}')\right] \mathcal{R}e\left[\mathcal{D}^*(\boldsymbol{\theta} - \boldsymbol{\theta}')\, g(\boldsymbol{\theta}')\right] \quad , \tag{2.8}$$

an equation which can be readily solved iteratively for $\kappa(\boldsymbol{\theta})$ for a specified data set $g(\boldsymbol{\theta})$.

In order to avoid the boundary effects which occur if (2.6) or (2.8) are applied to data in a finite region, we make use of the following relations, derived by Kaiser (1994): from appropriate combinations of the partial derivatives of the relations (2.3), one finds

$$\nabla \kappa(\boldsymbol{\theta}) = -\begin{pmatrix} \gamma_{1,1} + \gamma_{2,2} \\ \gamma_{2,1} - \gamma_{1,2} \end{pmatrix} \equiv \mathbf{U}(\boldsymbol{\theta}) \quad . \tag{2.9}$$

Hence, the gradient of $\kappa$ can be expressed in terms of the derivatives of the shear components, so that $\kappa(\boldsymbol{\theta})$ can be obtained as a line intergal

$$\kappa(\boldsymbol{\theta}) = \kappa(\boldsymbol{\theta}_0) + \int_{\boldsymbol{\theta}_0}^{\boldsymbol{\theta}} d\mathbf{l} \cdot \mathbf{U}(\mathbf{l}) \quad . \tag{2.10}$$

Hence, by any one choice of the integration path from $\boldsymbol{\theta}_0$ to all other points $\boldsymbol{\theta}$, one can in principle obtain $\kappa(\boldsymbol{\theta})$ at all points, up to an additive constant. However, it is clear that from noisy data, one can not recover a density field in this way. In order to reduce the noise in the reconstructed surface mass density one must average over many paths, i.e., use the information on the whole data field for each position $\boldsymbol{\theta}$, just as in the original inversion equation (2.6). We shall now give a prescription of how this averaging can be done.

Let the data field, on which $\gamma_i$ is given, have rectangular shape, of length $2L$ in the 1-direction, and $2rL$ in the 2-direction (most of what follows is in fact not restricted to a rectangular field, but this case will be most common in practice). Let $\mathbf{b}(\lambda)$, $0 \leq \lambda \leq \Lambda$, be a parametrization of the boundary curve of this data field, and let $\mathbf{l}_\lambda(t; \boldsymbol{\theta})$, $0 \leq t \leq 1$, be a curve which connects the point $\mathbf{b}(\lambda)$ with the point $\boldsymbol{\theta}$. Then, by averaging (2.10) over $\lambda$ [with $\boldsymbol{\theta}_0 = \mathbf{b}(\lambda)$], we obtain

$$\kappa(\boldsymbol{\theta}) = \frac{1}{\Lambda} \int_0^{\Lambda} \mathrm{d}\lambda \int_0^1 \mathrm{d}t \, \frac{\mathrm{d}\mathbf{l}_\lambda(t;\boldsymbol{\theta})}{\mathrm{d}t} \cdot \mathbf{U}(\mathbf{l}_\lambda(t;\boldsymbol{\theta})) + \frac{1}{\Lambda} \int_0^{\Lambda} \mathrm{d}\lambda \, \kappa(\mathbf{b}(\lambda)) \quad . \tag{2.11}$$

The important point to note is that the second term does not depend on $\boldsymbol{\theta}$, i.e., it is a constant. Since we already know that $\kappa$ can be determined only up to an additive constant, for a given distribution of the shear $\gamma$, eq. (2.11) does not introduce an additional uncertainty. Note that we could have included in the integral of (2.11) a properly normalized weight function of the form $w(\lambda)$, but this would be equivalent to a reparametrization of the boundary curve. Note that Eq. (2.11) is an exact inversion equation which makes use only of the shear on the finite data field.

The result of the integration (2.11) will depend on the choice of the curves connecting the boundary and the points $\boldsymbol{\theta}$, if noisy data are considered (for perfect data, the result is of course independent of this choice, but in this case, already (2.10) would be good enough). As was demonstrated in KS, their inversion formula (2.6) also is not unique, but optimal in the sense that it leads to the smallest errors – basically, (2.6) is the only inversion formula which does not single out a specific direction. We now want to rederive (2.6) from (2.11) and in this way find a hint for an appropriate choice of the curves $\mathbf{l}_\lambda$.

Hence, consider an isolated mass distribution (such that $\kappa$ vanishes outside a 'large' circle on the lens plane). For a point $\boldsymbol{\theta}$, define the curves

$$\mathbf{l}_\varphi(t;\boldsymbol{\theta}) = (1-t)R \begin{pmatrix} \cos\varphi \\ \sin\varphi \end{pmatrix} + \boldsymbol{\theta} \quad , \tag{2.12}$$

i.e., we use as parameter for the boundary curve the polar angle at the point $\boldsymbol{\theta}$. Note that the starting points ($t=0$) of the curves (2.12) depend on $\boldsymbol{\theta}$, but for a sufficiently large value of $R$, this is unimportant, since the additive term in (2.11) vanishes due to the assumption of an isolated mass distribution. One then has

$$\kappa(\boldsymbol{\theta}) = \frac{1}{2\pi} \int_0^{2\pi} \mathrm{d}\varphi \int_0^1 \mathrm{d}t \, \frac{\mathrm{d}\mathbf{l}_\varphi(t;\boldsymbol{\theta})}{\mathrm{d}t} \cdot \mathbf{U} \quad .$$

Since the curves cover the circle of radius $R$ exactly once, the integral can be transformed to Cartesian coordinates by defining $\boldsymbol{\theta}' = \mathbf{l}_\varphi(t;\boldsymbol{\theta})$; this yields

$$\kappa(\boldsymbol{\theta}) = \frac{1}{2\pi} \int \mathrm{d}^2\theta' \, \frac{\boldsymbol{\theta} - \boldsymbol{\theta}'}{|\boldsymbol{\theta} - \boldsymbol{\theta}'|^2} \cdot \mathbf{U}(\boldsymbol{\theta}') \quad . \tag{2.13}$$

By inserting $\mathbf{U}$ from (2.9), and integrating by part, one reobtains (2.6). The conclusion we can draw from this is that by an appropriate choice of the curves $\mathbf{l}_\lambda$, we can obtain a result as similar as possible to the inversion formula (2.6). The singularity at $\boldsymbol{\theta}' = \boldsymbol{\theta}$ of the integrand in (2.13) suggests that the curves should be chosen such that they locally correspond to the radial lines of a polar coordinate system centered on $\boldsymbol{\theta}'$; otherwise, this singular behaviour would amplify local noise in the shear data. On the other hand, we cannot choose simply radial parts, since the starting points of the curves have to be independent of $\boldsymbol{\theta}$, in order for the second term in (2.11) to be a constant.

Here, we shall take the following choice: the boundary curve of the rectangle is parametrized by the polar angle $\varphi$ as seen from the center of the rectangle. Then, for every point $\boldsymbol{\theta}$ inside the rectangle, we define a smaller rectangle of the same shape and orientation as the outer one. We define the size of this inner rectangle as follows: define $D_1 = \min(L - \theta_1, L + \theta_1)$, and $D_2 = \min(rL - \theta_2, rL + \theta_2)$; then, if $|\theta_2/\theta_1| \leq r$, the length of the rectangle in the 1-direction is chosen to be $2\eta D_1$, and otherwise $2\eta D_2/r$, and $\eta$ is a parameter between 0 and 1. We have illustrated this choice in Fig. 1a. Let $\mathbf{c}(\varphi; \boldsymbol{\theta})$ be the parametrization of the boundary curve of the small rectangle as given explicitly in the Appendix, where now $\varphi$ is the polar angle as measured from $\boldsymbol{\theta}$. Then, the curves $\mathbf{l}_\varphi$ are chosen such that, for each $\boldsymbol{\theta}$, it is a straight line from $\mathbf{b}(\varphi)$ to $\mathbf{c}(\varphi; \boldsymbol{\theta})$, and a radial line from $\mathbf{c}(\varphi; \boldsymbol{\theta})$ to $\boldsymbol{\theta}$ (for details, see the Appendix). In Fig. 1, these curves are drawn for several values of $\boldsymbol{\theta}$. With this choice, we have satisfied the two basic requirements: the starting points of the curves are independent of the value of $\boldsymbol{\theta}$, and they are distributed like the radial coordinate lines near $\boldsymbol{\theta}$. Of course, this choice is still largely arbitrary, and one cannot expect that it yields an 'optimal' reconstruction of the surface mass density. But as shall be demonstrated below, it removes the systemmatic boundary effects inherent in applying (2.6) to a finite data field.

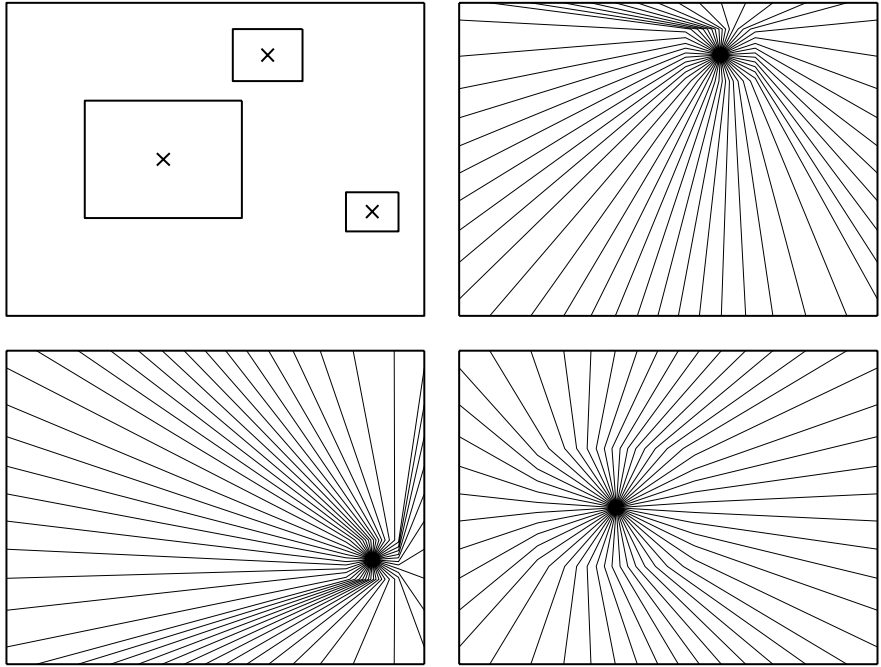

**Fig. 1.** The choice of our integration curves $\mathbf{l}_\varphi(t; \boldsymbol{\theta})$. In the upper left panel, the boundary curves $\mathbf{c}(\varphi, \boldsymbol{\theta})$ of the inner rectangles for three different points $\boldsymbol{\theta}$ (marked as crosses) are shown, for $\eta = 0.5$ which is taken throughout this paper. The three other panels show the curves $\mathbf{l}_\varphi$ according to these three points $\boldsymbol{\theta}$

As mentioned before, the shear $\gamma$ is not an observable, but the quantity $g$ (2.7) can be measured locally if the cluster is noncritical. By inserting the definition (2.7) into (2.9), one obtains (see Kaiser 1994)

$$\nabla K(\boldsymbol{\theta}) = \frac{1}{1 - g_1^2 - g_2^2} \begin{pmatrix} 1 + g_1 & g_2 \\ g_2 & 1 - g_1 \end{pmatrix} \begin{pmatrix} g_{1,1} + g_{2,2} \\ g_{2,1} - g_{1,2} \end{pmatrix} \equiv \mathbf{u}(\boldsymbol{\theta}) \quad, \tag{2.14}$$

where
$$K(\boldsymbol{\theta}) := \ln(1 - \kappa(\boldsymbol{\theta})) \quad . \tag{2.15}$$

Hence, it is possible to derive the gradient of the quantity $K$ in terms of the observable quantity $g$, and the integration method for (2.15) can be chosen in the same way as for (2.9), i.e.,

$$K(\boldsymbol{\theta}) = \frac{1}{2\pi} \int_0^{2\pi} \mathrm{d}\varphi \int_0^1 \mathrm{d}t \, \frac{\mathrm{d}\mathbf{l}_\varphi(t;\boldsymbol{\theta})}{\mathrm{d}t} \cdot \mathbf{u}(\mathbf{l}_\varphi(t;\boldsymbol{\theta})) + \frac{1}{2\pi} \int_0^{2\pi} \mathrm{d}\varphi \, K(\mathbf{b}(\varphi)) \quad . \tag{2.16}$$

Of course, $K$ can only be determined up to an additive constant, or $(1-\kappa)$ can only be determined up to a constant factor, which expresses the fact (noted also by Kaiser 1994) that there is a global invariance transformation of the surface mass density which leaves the observable distortions unchanged (see Sect. 3.4 of Paper I).

## 3 Application to synthetic data

As in Papers I & II, we generate distorted images of background sources by distributing galaxies randomly onto the lens plane (this is an approximately valid procedure, since the local slope of the source counts of faint galaxies is such that the decrease of the number density of galaxy images due to the solid angle distortion by the light deflection is compensated by the magnification bias; in other words, the local slope of the cumulative source counts is close to $-1$, in which case the number counts are unchanged by lensing – see Sect. 12.1.1 of SEF). For each galaxy, we draw an intrinsic ellipticity from an assumed ellipticity distribution, assuming that the intrinsic orientation of the sources are distributed randomly; note that this basic assumption lies at the heart of all reconstruction methods. Then, for a chosen surface mass distribution (which we want to reconstruct), the local lensing parameters, i.e., surface mass density and shear, are calculated at the position of each galaxy, and the ellipticity of the lensed galaxy image is calculated from the intrinsic ellipticity and the lens parameters. This set of 'observed' images is then used to reconstruct the mass density of the lens.

In Papers I & II, we have considered the (complex) ellipticity

$$\chi = \frac{Q_{11} - Q_{22} + 2\mathrm{i}Q_{12}}{Q_{11} + Q_{22}} \tag{3.1}$$

in terms of the tensor $Q$ of second brightness moments of an image (and a similar definition applies for the intrinsic ellipticity of the source). The local lens equation then yields the transformation between the source and image ellipticity as

$$\chi^{(\mathrm{s})} = \frac{2g + \chi + g^2 \chi^*}{1 + |g|^2 + 2\mathcal{R}\mathrm{e}\,(g\chi^*)} \quad , \tag{3.2}$$

where $g$ is given by (2.7). If $R \in [0,1]$ denotes the ratio of the moduli of the eigenvalues of the moment tensor $Q$, then the modulus of $\chi$ is $|\chi| = (1-R^2)/(1+R^2)$. Here we want to use a somewhat different ellipticity parameter $\epsilon$, previously used also by other authors (e.g., Bonnet & Mellier 1994, Schramm & Kayser 1994); the phase of $\epsilon$ is defined to be the same as that of $\chi$, and its modulus in terms of the ratio $R$ of the moduli of the eigenvalues of the moment tensor $Q$ is $|\epsilon| = (1 - R)/(1 + R)$. This leads to the relation between $\chi$ and $\epsilon$:

$$\chi = \frac{2\epsilon}{1 + |\epsilon|^2} \quad ; \quad \epsilon = \frac{\chi}{1 - \sqrt{1 - |\chi|^2}} \quad ; \quad (3.3)$$

analogous relations hold for the ellipticity of the sources. Since we consider here only noncritical clusters, the transformation between $\epsilon$ and $\epsilon^{(s)}$ is unique and obtained by inserting (3.3) into (3.2),

$$\epsilon^{(s)} = \frac{g + \epsilon}{1 + g^*\epsilon} \quad ; \quad \epsilon = \frac{\epsilon^{(s)} - g}{1 - g^*\epsilon^{(s)}} \quad . \quad (3.4)$$

In the critical region of clusters (i.e., where the determinant of the matrix $A$ (2.2) is negative), (3.4) is no longer valid and has to be replaced by somewhat different relations, whereas (3.2) is true in general. Hence, the quantity $\epsilon$ is particularly convenient only in the noncritical case considered here. The reason why we here prefer to work in terms of $\epsilon$ instead of $\chi$ is the convenient property of the mean of $\epsilon$ over a set of images,

$$\langle \epsilon \rangle = -g \quad , \quad (3.5)$$

independent of the intrinsic ellipticity distribution (Schramm & Kayser 1994) as long as the intrinsic orientation of the sources are randomly distibuted. The property (3.5) can be easily checked by angular integration of the relations (3.4). Further investigations of the statistical properties of $\epsilon$ will be published elsewhere.

Hence, (3.5) can be conveniently used to determine $g$ locally, by averaging over a number of galaxies at each position. Specifically, to determine an estimate for $g(\boldsymbol{\theta})$, the same averaging procedure as in Paper II is used,

$$g(\boldsymbol{\theta}) = -\frac{\sum_i w_i \epsilon_i}{\sum_i w_i} \quad , \quad (3.6)$$

with weights

$$w_i \propto \exp\left(\frac{(\boldsymbol{\theta} - \boldsymbol{\theta}_i)^2}{\Delta\theta^2}\right) \quad , \quad (3.7)$$

and the smoothing scale $\Delta\theta$ can be chosen appropriately. As in Paper II, the smoothing scale is adopted to the 'strength' of the signal, i.e., smaller smoothing scales are employed in regions of larger shear. For more details, see Paper II. In this way we have calculated $g$ on a grid in the lens plane, and obtained the partial derivatives of the components of $g$ by finite differencing. Hence, the vector $\mathbf{u}$ also has been calculated on a grid. The $t$-integral in (2.16) was then performed by bilinear interpolating on the grid. The iteration in (2.8) converges after a few steps.

Another smoothing is introduced in the application of (2.6): in order to avoid the singular denominator, we have multiplied the integrand in (2.6) by a factor $W(|\boldsymbol{\theta} - \boldsymbol{\theta}'|)$, where

$$W(x) = 1 - \left(1 + \frac{x^2}{2s^2}\right) \exp\left(-\frac{x^2}{2s^2}\right) \quad (3.8)$$

as in Paper II. In all the applications shown below, the smoothing introduced by (3.6) is much more important than the introduction of the factor $W$, since we choose $s \ll \Delta\theta$.

For our illustrative calculations, we have assumed an intrinsic ellipticity distribution of the form

$$p_\mathrm{s}(\epsilon^{(\mathrm{s})}) = \frac{1}{\pi \rho^2 (1 - \mathrm{e}^{-1/\rho^2})} \mathrm{e}^{-|\epsilon^{(\mathrm{s})}|^2/\rho^2} \quad , \tag{3.9}$$

where $p_\mathrm{s}(\epsilon^{(\mathrm{s})}) \, \mathrm{d}^2\epsilon^{(\mathrm{s})}$ is the probability that the source ellipticity lies within $\mathrm{d}^2\epsilon^{(\mathrm{s})}$ of $\epsilon^{(\mathrm{s})}$. Hence, the quantity $\rho$ controls the width of the intrinsic ellipticity distribution, and we expect that with increasing $\rho$, the reconstructed mass density will become noisier. In all cases presented below, we have fixed the source density to be $40/(\mathrm{arcmin})^2$. In all cases, the field on which galaxy images are distributed is slightly larger than the field shown in the figures.

## 4 Examples

In this section, we consider several examples of mass reconstructions, which were performed by the methods described in the preceding sections. As was remarked earlier, these reconstruction methods yield $\ln(1 - \kappa)$ only up to an additive constant. Therefore, in all figures which follow we plot contours of constant $\kappa$, which are spaced by 0.02 in $K = \ln(1 - \kappa)$.

As a first example, we consider a single isothermal sphere, with data in a square-shaped field of sidelength 10 arcmin. The surface mass density is shown in Fig. 2a. The lens was chosen to have a core radius of 1 arcmin, and a central surface mass density of 0.8. In Fig. 2b we have plotted the result from the inversion according to (2.8), where we have used the exact values for $g$ at each position. One can see that this method leads to systemmatic boundary effects, namely local minima close to the sides of the square, saddle points on the diagonals, and an increase of $\kappa$ towards the corners. In addition, the contours deviate more and more from their circular shape as the boundaries are approached. As was mentioned before, these artefacts are due to the finite region of the lens plane over which the integration in (2.8) is performed. If we use the exact data for $g$ in (2.16), we would reobtain the original mass distribution.

In Figs. 2c&d two reconstructions according to (2.8) and (2.16) are shown, respectively. Here we use a galaxy density of $40/(\mathrm{arcmin})^2$, an ellipticity distribution of the sources according to (3.9), with $\rho = 0.2$, and a smoothing scale of $\Delta\theta = 1$ arcmin in the regions of weak distortions, with smaller smoothing scales where the distortion signal becomes larger – see Paper II. By comparing the two reconstructions, we first note that their 'noise levels' are nearly identical. Hence, at first sight the quality of the reconstructions from (2.8) and (2.16) is the same. However, a closer look then shows that the systemmatics, visible in Fig. 2b, are also present in Fig. 2c: despite the noise, caused by the discreteness of the galaxy images and their intrinsic ellipticity distibution, one can still recognize the local minima near the sides of the square and the rise of $\kappa$ towards the corners. These features are not seen in Fig. 2d – though detailed features in the maps which are due to the realization of the galaxy distribution can be linked to each other, there seem to be fewer systematic structures in the reconstruction according to (2.16). However, in the example shown in Fig. 2, the systemmatic boundary effects are at a fairly low level, affecting only contour levels which are already fairly noisy due to the noise caused by the discreteness of the galaxy images and the intrinsic ellipticity distribution. On the other hand, the situation in Fig. 2 is most favourable for the inversion

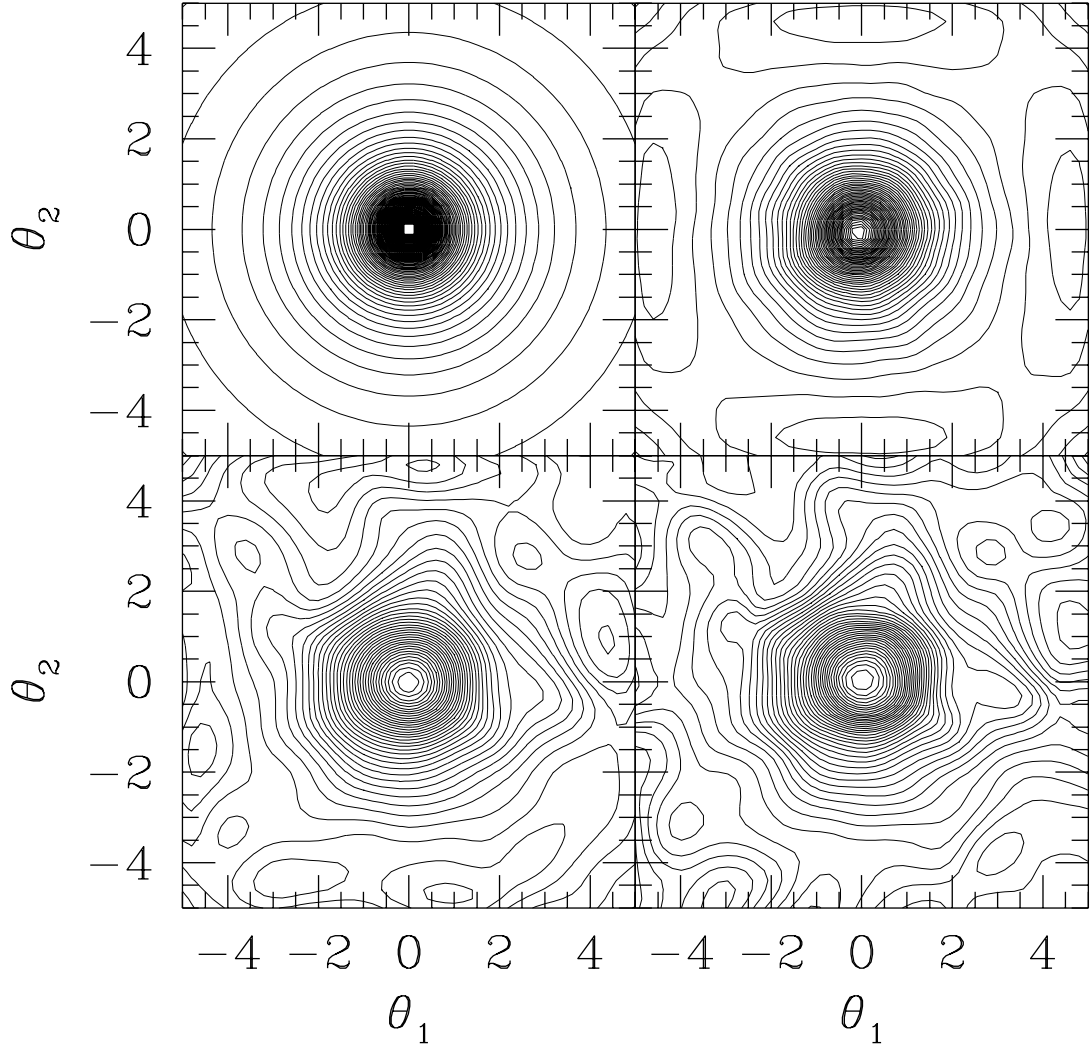

**Fig. 2.** Surface mass density reconstruction for a single isothermal sphere with finite core. The lens is at the center of each frame, which is a square of 10 arcmin size. The lines plotted are contours of constant $\kappa$, and they are scaled such that the spacing between two adjecent contours is 0.02 in $\ln(1-\kappa)$. The core radius of the lens is 1 arcmin. (a)–upper left panel: the original mass distribution. (b)–upper right panel: reconstruction with the KS method, i.e., using (2.8), with 'perfect data', i.e., without noise due to the discreteness of galaxy images and their ellipticity distribution; note that a reconstruction according to (2.16) with perfect data would yield the original mass distribution shown in (a). The two lower panels are true reconstructions, with a galaxy density of $40/(\mathrm{arcmin})^2$, and an ellipticity distribution of the form (3.9), with $\rho = 0.2$. The smoothing scale $\Delta\theta$ in (3.7) was chosen to be 1 arcmin in regions of low distortions, and decreased as the distortions become larger – cf. Paper II. (c)–lower left panel: Reconstruction according to (2.8). (d)–lower right panel: reconstruction using (2.16)

formula (2.8), since it contains an isolated matter distribution centered on a fairly large data field.

To see the systemmatic boundary effects more clearly, we have plotted in Fig. 3 the analogous reconstruction with a rectangular data field with axis ratio $r = 2/3$. In Fig. 3b, the boundary effects are obvious, with minima close to the sides of the rectangle, saddle points on the diagonals, and again with $\kappa$ increasing towards the corners. These features are also seen in the reconstruction performed with (2.8), Fig. 3c, whereas they are largely absent in the reconstruction in Fig. 3d, performed with (2.16). Again, the noise level in the two reconstructions is about the same, but the one in panel (d) lacks the systemmatic effects.

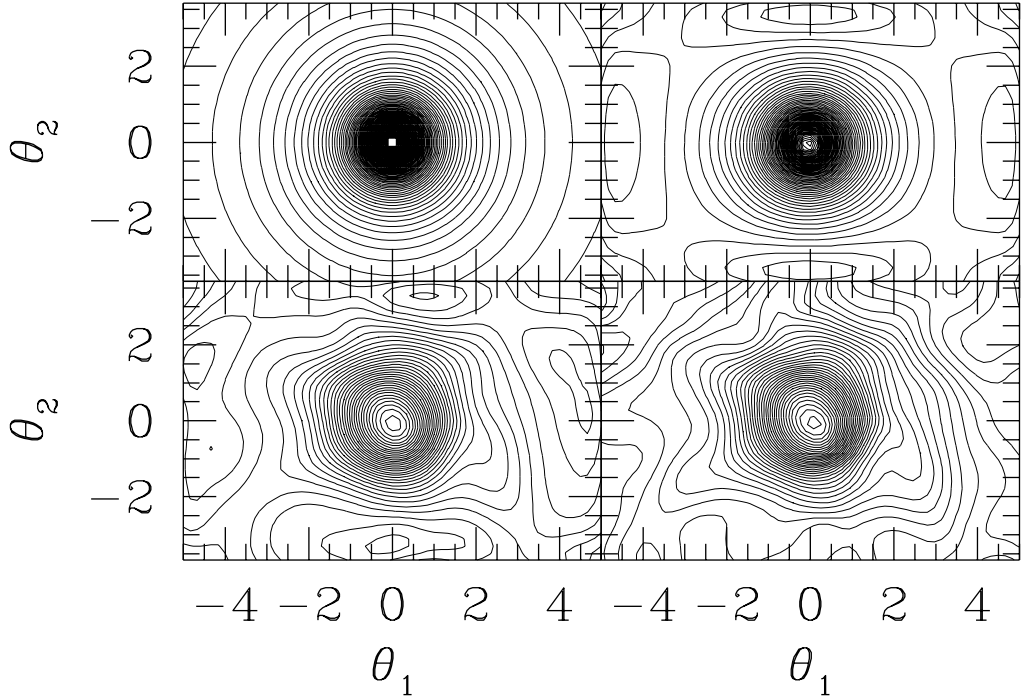

**Fig. 3.** Same as in Fig. 2, but the data field is now rectangular, with length 10 arcmin, and side ratio of $r = 2/3$

    We give two further illustrations in Figs. 4 & 5. In Fig. 4, we have taken the same lens model as in the previous two figures, but choose the size of the data field to be 6 arcmin, and the lens is placed close to the boundary of the frame. Fig. 4b shows that, as expected, the systematic effects increase relative to the case that the lens was centered on the field. In particular, there is a broad, very flat 'plateau' in the $\kappa$ distribution, and minima

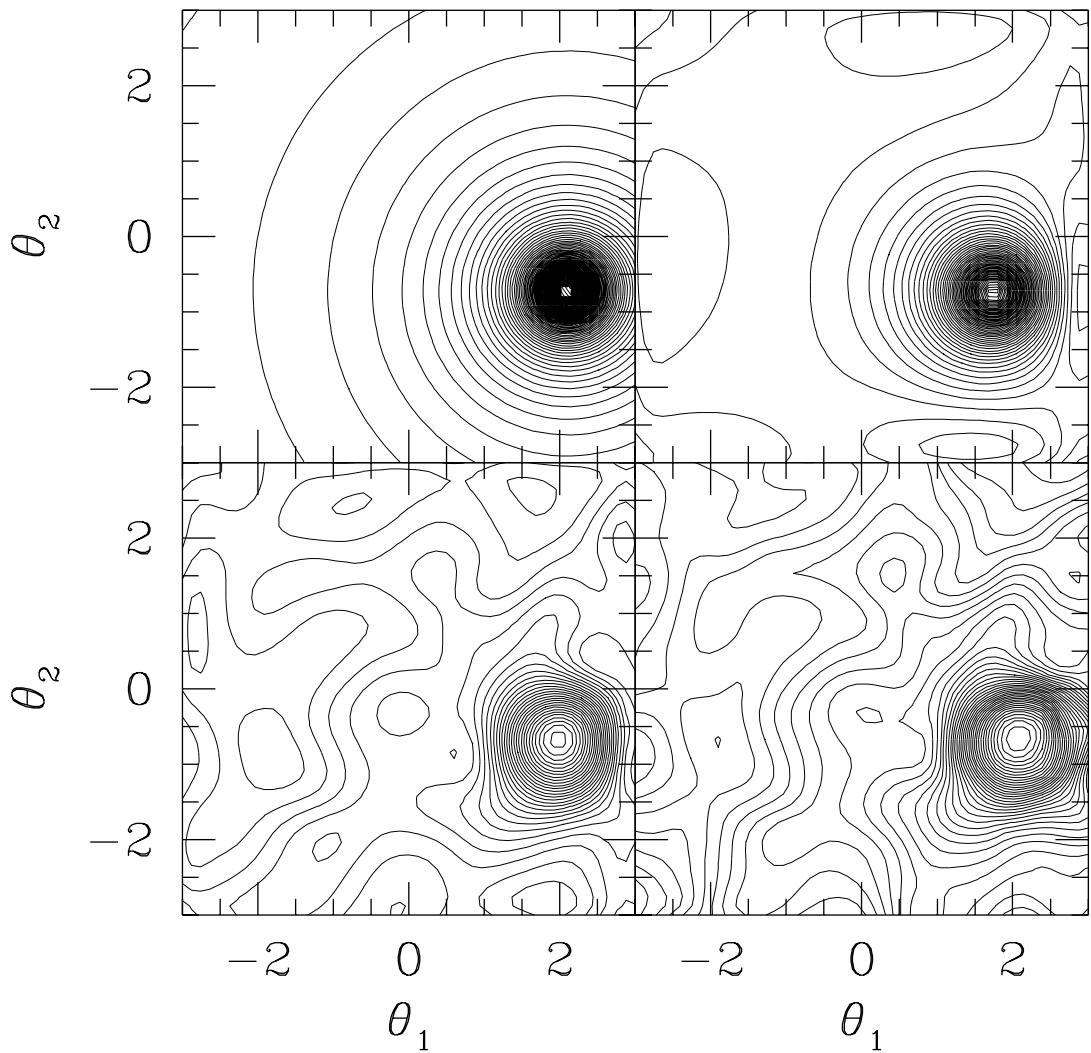

**Fig. 4.** Same as Fig. 2, but now the data field is a square of sidelength 6 arcmin, and the isothermal sphere is not centered on the field, but close to one of its edges

between the center of the lens and the nearest boundaries. These artefacts remain visible in the reconstruction, Fig. 4c, whereas they are basically absent in the reconstruction shown in Fig. 4d. Finally, in Fig. 5 we have plotted the reconstruction of a lens consisting of two spherical components. The systemmatics caused by the inversion (2.8) are most clearly seen in Fig. 5b, with pronounced minima between the left lens component and the boundary; these features survive nearly unchanged in the reconstruction shown in Fig. 5c, whereas they are absolutely absent in the reconstruction performed with (2.16), as can be seen in Fig. 5d. To see this more clearly, the same data as in Fig. 5c,d are plotted in Fig. 6. Here it can be seen that the boundary effects are quite dramatic. In addition, this figure clearly shows that the 'amplitude' in the variation of $K = \ln(1 - \kappa)$ over the field is underestimated by the inversion formula (2.8), probably also because the integration is truncated beyond the data field. The reason why the contours in Fig. 5 are much smoother than in the other cases is the strength of the lens here: first, there are two lenses causing the distortions, instead of one, and in addition, the shear between the two lenses is particularly strong, whereas an isolated isothermal sphere does not lead to strong distortions, as long as it is noncritical.

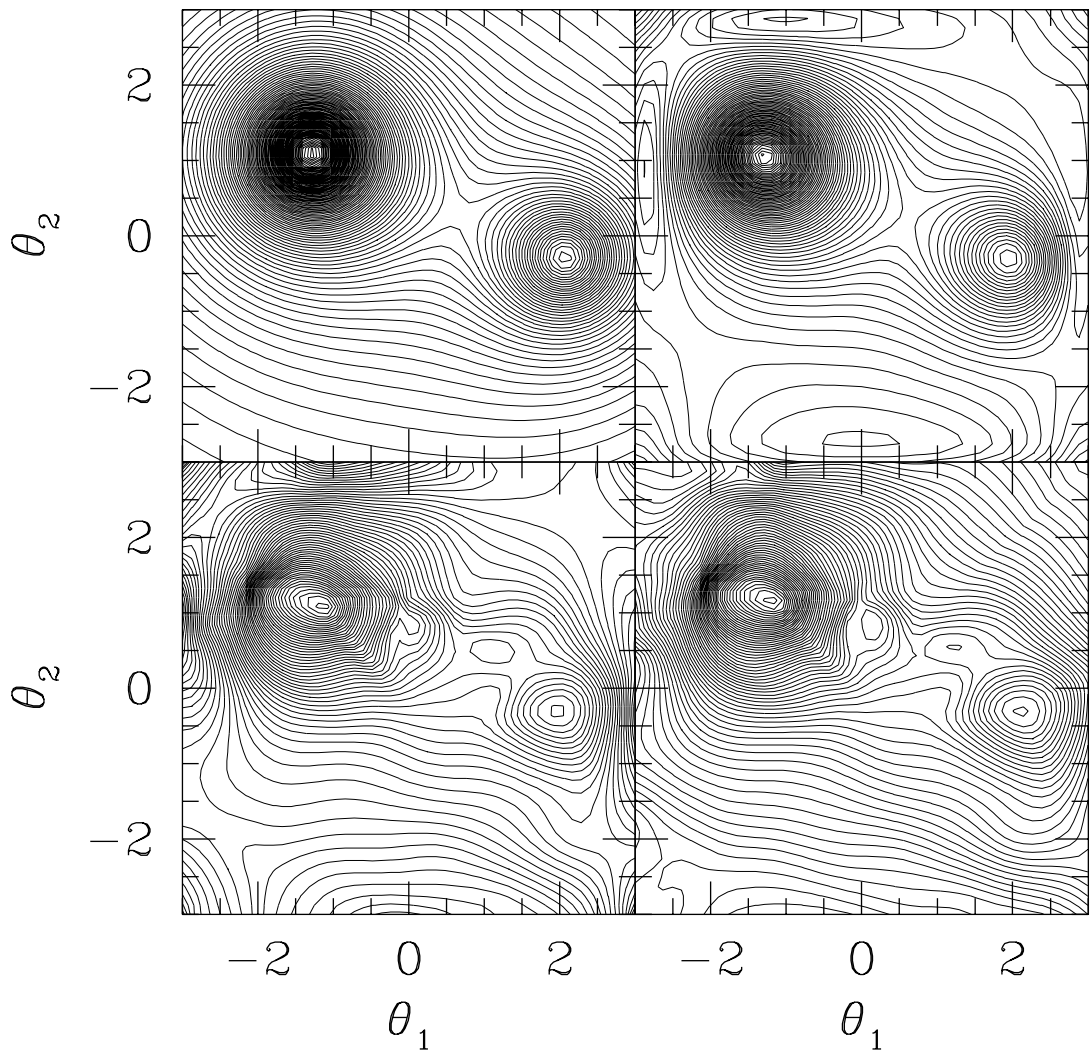

**Fig. 5.** Same as Fig. 4, but now the lens consists of two components, each one modelled as an isothermal sphere

From the examples shown above, we can conclude that the new inversion formula (2.16), which is based on the differential equation (2.14) derived by Kaiser (1994), is indeed useful: it lacks the systemmatic effects with which (2.8) is burdened, and the noise level of the reconstructed surface mass density is comparable to that obtained by (2.8). Since the inversion formula (2.16) explicitly is constrained to data inside the data field, it is not expected to show any systemmatic boundary effects; however, this does not exclude that the noise level increases towards the boundaries of the field.

# 5 Summary and discussion

In this paper we have investigated a new method to derive the surface mass density of a lens (e.g., a cluster) from the distortion of the images of background sources (faint galaxies). This cluster inversion problem has been solved previously by Kaiser & Squires (1993); they have derived an inversion equation which is exact if the distortion data are available over the whole lens plane, and is a very useful approximation if the data field extends over most of the lensing region. However, due the small size of most currently used CCDs, this latter condition is not always satisfied if the clusters have a large angular extent. It is therefore desireable to have an inversion formula which explicitly makes use only of data in a finite field (i.e., the CCD). A recently published result by Kaiser (1994) has been the starting point of the current investigation; he expressed the gradient of the quantity $K = \ln(1-\kappa)$ in terms of observables (at least in the case of noncritical lenses, i.e., lenses which are not capable of producing multiple images). By an appropriate integration of that gradient, we have derived an explicit formula for $K(\boldsymbol{\theta})$ which uses only the distortion within a finite data field. The resulting equation therefore is exact on a finite region. The surprising result, demonstrated in Sect. 4, that the 'noise' of this new inversion formula is not appreciably larger than that of the KS inversion (2.8) makes this new method useful, since it is free of systemmatic effects which are inherent in applying (2.8) to a finite data field. We have provided several illustrations of these systemmatic effects as a warning against careless application of (2.8). In particular, we have shown in Fig. 3 that the results from the inversion formula (2.8) should be interpreted with great care if a rectangular CCD with side ratio not close to unity is used. This point was already stressed in Paper II; we presently consider the results of the inversion of the cluster 0016+16 shown in Fig. 7 of Smail et al. (1994) as not reliable. A comparison of this figure with our Fig. 3b,c shows that their features at both ends of the rectangle may simply be artefacts of the geometry of the data field and the properties of (2.8). In order to check the validity of this assertion, it would be useful to apply our new inversion formula (2.16) to the data field of 0016+16. We also want to note that there is no fundamental difficulty to generalize (2.16) to the case of critical clusters, basically using the same procedure as in Paper II.

Though we believe that our inversion formula (2.16) is in some sense superiour to (2.8), we are convinced that the present formula is not the best one can obtain. We have described a set of curves $\mathbf{l}_\varphi$ which can be used in (2.16); however, the choice of this set of curves was to a large degree arbitrary, except that we had to satisfied two constraints which were discussed in Sect. 2. Note that the inversion formula (2.8) is 'optimal' on $\mathbb{R}^2$. It is therefore not surprising that the 'noise level' in some of our examples is slightly larger if the reconstruction formula (2.16) is used than that from (2.8). As was already remarked by Kaiser (1994), one could construct an inversion formula from (2.14) even if some parts of the data field cannot be used (e.g., because a bright foreground galaxy outshines the faint background images), by choosing curves $\mathbf{l}$ which avoid these unusable regions, although it probably will be difficult to construct these curves in a way to 'minimize' the noise in the resulting reconstruction.

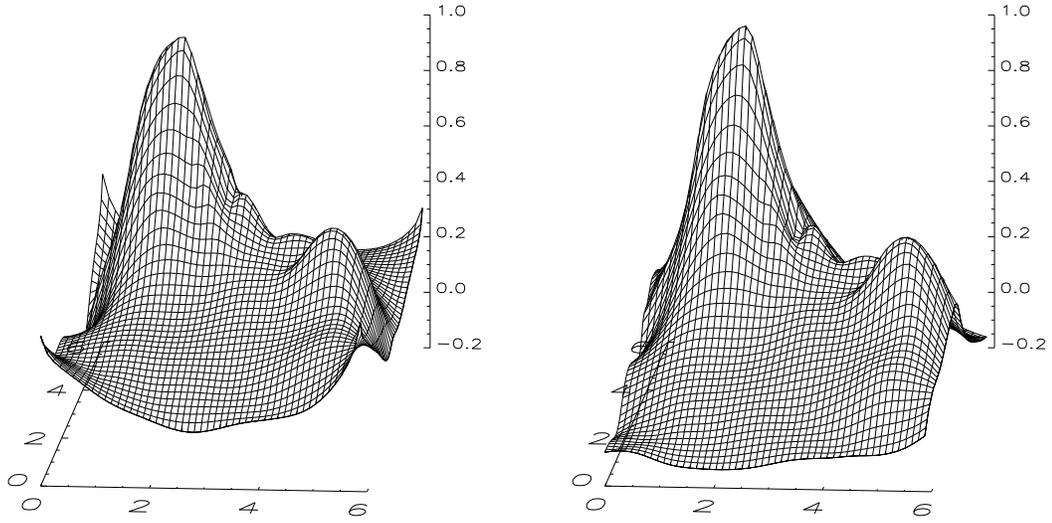

**Fig. 6.** The same reconstructions as in Fig. 5, in a different graphical representation. Left and right panel correspond to the reconstruction according to (2.8) and (2.16), respectively. This representation shows more clearly the boundary effects in the reconstruction by (2.8), i.e., the rising of $K$ towards the corners, and the deep minimum of $K$ between the major peak and the boundary. These features are basically absent in the reconstruction according to (2.16), but one also sees that the reconstruction according to (2.8) yields a somewhat smoother result. In particular, the reconstruction according to (2.8) is somewhat better in the central part of the field

It remains an open problem whether there exists also an 'optimal' inversion equation on a finite field, how this equation looks like, and 'how far' our equation (2.16) is away from such an optimal inversion formula. Since the observations of faint distortions requires great efforts, manpower and costs, the development of the best theoretical tool to analyze the observational data is certainly justified and necessary.

I would like to thank J. Ehlers, H.-W. Rix, C. Seitz, S. Seitz, J. Wambsganss and A. Weiss for useful discussions.

## Appendix

In this appendix, we give explicit equations for the curves $\mathbf{l}_\varphi(t; \boldsymbol{\theta})$ which appear in the integral (2.16) and which are described in Sect. 2. Define $\Phi_0 = \arctan r$, where $r$ is the side ratio of the rectangle, and the four $\varphi$-intervals $I_1 = [-\Phi_0, \Phi_0]$, $I_2 = [\Phi_0, \pi - \Phi_0]$, $I_3 = [\pi - \Phi_0, \pi + \Phi_0]$, $I_4 = [\pi + \Phi_0, 2\pi - \Phi_0]$. The length of the inner rectangle around a point $\boldsymbol{\theta} = |\boldsymbol{\theta}| (\cos\vartheta, \sin\vartheta)$ is $2a(\boldsymbol{\theta})$, where

$$a(\boldsymbol{\theta}) = \eta \begin{cases} L - \theta_1 & \text{for } \vartheta \in I_1 \\ L - \theta_2/r & \text{for } \vartheta \in I_2 \\ L + \theta_1 & \text{for } \vartheta \in I_3 \\ L + \theta_2/r & \text{for } \vartheta \in I_4 \end{cases}, \qquad (A1)$$

and $\eta \in [0, 1]$ can be chosen appropriately. Throughout the paper, we used $\eta = 1/2$. The parametrization of the boundary curve $\mathbf{b}(\varphi)$ of the data field is

$$\mathbf{b}(\varphi) = \begin{cases} L\mathbf{s}(\varphi) & \text{for } \varphi \in I_1 \\ rL\boldsymbol{\sigma}(\varphi) & \text{for } \varphi \in I_2 \\ -L\mathbf{s}(\varphi) & \text{for } \varphi \in I_3 \\ -rL\boldsymbol{\sigma}(\varphi) & \text{for } \varphi \in I_4 \end{cases}, \qquad (A2)$$

where
$$\mathbf{s}(\varphi) = \begin{pmatrix} 1 \\ \tan\varphi \end{pmatrix} \quad ; \quad \boldsymbol{\sigma}(\varphi) = \begin{pmatrix} \cot\varphi \\ 1 \end{pmatrix} \quad . \tag{A3}$$

The boundary curve $\mathbf{c}(\varphi;\boldsymbol{\theta})$ of the inner rectangle around $\boldsymbol{\theta}$ is then given by

$$\mathbf{c}(\varphi;\boldsymbol{\theta}) = \boldsymbol{\theta} + \begin{cases} a(\boldsymbol{\theta})\mathbf{s}(\varphi) & \text{for } \varphi \in I_1 \\ ra(\boldsymbol{\theta})\boldsymbol{\sigma}(\varphi) & \text{for } \varphi \in I_2 \\ -a(\boldsymbol{\theta})\mathbf{s}(\varphi) & \text{for } \varphi \in I_3 \\ -ra(\boldsymbol{\theta})\boldsymbol{\sigma}(\varphi) & \text{for } \varphi \in I_4 \end{cases} \quad . \tag{A4}$$

Then, the curves $\mathbf{l}_\varphi(t;\boldsymbol{\theta})$ which appear in (2.16) are given by

$$\mathbf{l}_\varphi(t;\boldsymbol{\theta}) = \begin{cases} (1-2t)\mathbf{b}(\varphi) + 2t\mathbf{c}(\varphi;\boldsymbol{\theta}) & \text{for } t \in [0,1/2] \\ (2-2t)\mathbf{c}(\varphi;\boldsymbol{\theta}) + (2t-1)\boldsymbol{\theta} & \text{for } t \in [1/2,1] \end{cases} \quad . \tag{A5}$$